%% file: main.tex
\documentclass[10pt,conference]{IEEEtran}
\IEEEoverridecommandlockouts
\usepackage{cite}
\usepackage{amsmath,amssymb,amsfonts}
\usepackage{algorithmic}
\usepackage{graphicx}
\usepackage{textcomp}
\usepackage{xcolor}
\def\BibTeX{{\rm B\kern-.05em{\sc i\kern-.025em b}\kern-.08em
    T\kern-.1667em\lower.7ex\hbox{E}\kern-.125emX}}

\usepackage{xspace}
\usepackage{subcaption} 
\usepackage{multirow}
\usepackage{booktabs}

\usepackage{hyperref} 
\hypersetup{hidelinks}
\usepackage{url}

\usepackage{makecell} 
\usepackage{tcolorbox}

\usepackage[normalem]{ulem}

\usepackage{amsfonts,amssymb}

\newcommand{\toolname}[0]{{\emph{CodeMark}}\xspace} 
\newcommand{\awt}[0]{{\emph{AWT}}\xspace}
\newcommand{\awtcode}[0]{{$AWT_{code}$}\xspace}
\newcommand{\cals}[0]{{\emph{CALS}}\xspace}
\newcommand{\calscode}[0]{{$CALS_{code}$}\xspace}

\newcommand{\blueDelete}[1]{}
\newcommand{\blue}[1]{{#1}}

\newcommand{\eg}[0]{{\em e.g., }}
\newcommand{\etal}{\hbox{\emph{et al.}}\xspace}

\begin{document}

\title{Towards Tracing Code Provenance \\ with Code Watermarking}

\author{
  \IEEEauthorblockN{Wei Li, Borui Yang, Yujie Sun, Suyu Chen, \\ Ziyun Song, Liyao Xiang, Xinbing Wang}
  \IEEEauthorblockA{\textit{Shanghai Jiao Tong University}\\
  \{li\_wei, ybirua, sunyujie, csy\_ichigo, aliceziyun, xiangliyao08, xwang8\}@sjtu.edu.cn }
  \and
  \IEEEauthorblockN{Chenghu Zhou}
  \IEEEauthorblockA{Chinese Academy of Sciences \\  
  zhouchsjtu@gmail.com }
}

\maketitle

\input{sections/0_abstract}


\input{sections/1_introduction}

\input{sections/7_related}

\input{sections/3_approach}
\input{sections/4_experiment_set}

\input{sections/5_results}
\input{sections/6_discussion}
\input{sections/8_conclusion}

\bibliographystyle{IEEEtran}
\bibliography{references}

\end{document}

%% file: sections/0_abstract.tex
\begin{abstract}
Recent advances in large language models have raised wide concern in generating abundant plausible source code without scrutiny, and thus tracing the provenance of code emerges as a critical issue. To solve the issue, we propose \toolname, a watermarking system that hides bit strings into variables respecting the natural and operational semantics of the code. For naturalness, we novelly introduce a contextual watermarking scheme to generate watermarked variables more coherent in the context atop graph neural networks. Each variable is treated as a node on the graph and the node feature gathers neighborhood (context) information through learning. Watermarks embedded into the features are thus reflected not only by the variables but also by the local contexts. We further introduce a pre-trained model on source code as a teacher to guide more natural variable generation. Throughout the embedding, the operational semantics are preserved as only variable names are altered. Beyond guaranteeing code-specific properties, \toolname is superior in watermarking accuracy, capacity, and efficiency due to a more diversified pattern generated. Experimental results show \toolname outperforms the SOTA watermarking systems with a better balance of the watermarking requirements.
\end{abstract}

%% file: sections/1_introduction.tex
\section{Introduction}

Tracing the provenance of source code snippets (e.g., functions) is a crucial yet under-exploited topic \cite{liu2021can, li2022ropgen}. With the recent breakthrough in large language models (LLMs) \cite{ChatGPT, GPT4, palm}, the topic has raised wide concern due to the potential misuse of LLMs in distributing plausible, fake, or even malicious code snippets \cite{dakhel2022github, banchatgpt} without close inspection. Yet it is difficult to examine the code as LLM service providers usually do not retain its generated code. It is even harder to inspect human-written code for proprietary or privacy reasons. Hence many code provenance techniques including clone detection \cite{liu2021can, eghbali2022crystalbleu, sajnani2016sourcerercc, zhang2019novel} and authorship recognition \cite{ou2022scs, li2022ropgen, liu2021practical, abuhamad2018large} have been hindered in these circumstances.

Watermarking has shown to be a promising technique in tracing image/text/software provenance \cite{zhu2018hidden, abdelnabi2021adversarial, kadian2021robust, dey2019software}. A watermarking system typically consists of an embedding module and an extraction module \cite{ma2019xmark}. Provided with the original subject and a bit string, the embedding module generates a new subject with high similarity to the original but hides the bit string as a unique identifier of the owner. The extraction module, fed with the watermarked subject, produces the watermark as owner-proof.

Despite prosperous results in watermarking, there are few works on source code watermarking. Source code is one of a kind, which has distinctive features from natural languages: it involves two channels of information --- formal and natural \cite{chakraborty2022natgen, casalnuovo2020theory}. The formal channel affords operational semantics used by interpreters, compilers, etc. while the natural channel is commonly used by humans for code comprehension and communication. While inconspicuous words and symbols (such as articles and semicolons) changes in texts would not increase understanding difficulty, they may be detrimental to code. For instance, a misplacement of a semicolon in Java yields code that fails to run, or a substitute variable has a misleading name to confuse developers \cite{jiang2022bugs}. In fact, $21.7\%$ of patches in Eclipse and Mozilla projects were rejected because the patches contain bad identifier names \cite{tao2014writing}. Traditional software watermarking methods embed bit strings into obfuscated or encrypted executable files \cite{el2004hydan, collberg2005software, dey2019software}, which destroy the natural semantics and thus is unacceptable for source code. 

\begin{figure}[htbp]
	\centerline{\includegraphics[scale=0.5]{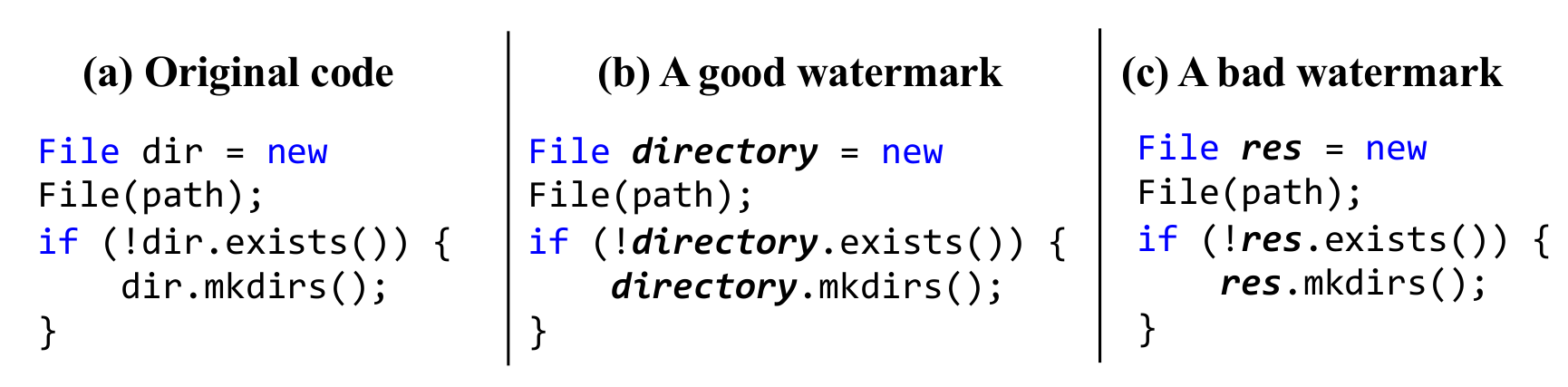}}
	\caption{A good and a bad example of code watermarking.}
	\label{fig:goodbad}
\end{figure}

To address the issue, we propose \toolname, a code watermarking system as a solution for automatically hiding bit strings in source code considering the dual-channel property. The solution takes advantage of the ample space for variable names and applies machine translation-like techniques to generate substitute variables both meaningful in the original context while hiding watermarks. Since the formal channel of the code is unaltered, the operational semantics remain unchanged. To generate watermarked code as natural as the original, we novelly propose \textit{contextual watermarking} --- embedding watermarks to variable names according to the context where the variable is located. The context, which could be described by the abstract syntax tree of the code, reflects the unique, structural characteristics of the code. Take the code snippets in Fig.~\ref{fig:goodbad} as an example. The original code creates a new file directory and variable \texttt{dir} is the target variable. Replacing it with the word \texttt{directory} is reasonable and easy to understand, while substitution \texttt{res} would be confusing to developers.

Taking the code context into account, we model the code by graphs and variables as nodes on graphs. A graph neural network (GNN) is adopted in both embedding and extraction modules to learn the contextual feature of the target variable. Graph learning essentially aggregates the neighborhood information as features of the target node. Watermark is embedded by acting as a tunable knob selecting features from the output of the multi-head attention layers of GNN. Different from fixed-rule substitution generating rigid patterns, we observe our watermark patterns are more diversified, e.g., the same variable substitution in varied contexts could refer to different bit strings. To further enhance the naturalness of the code, we utilize a pre-trained model on source code as a teacher to distill the generic representation of programming language to our embedding module. We train our embedding and extraction modules end to end w.r.t. the watermark loss and the naturalness loss.

Highlights of our contributions are: we build a contextual and natural watermarking system for code, \toolname, which is operational semantics-preserving while incurring little naturalness loss. Experiments show that our system outperforms the SOTA natural language watermarking tools and their variants for code in terms of a balance of watermarking requirements, i.e., effectiveness, secrecy, accuracy, capacity, and robustness. We will publish our source code.

%% file: sections/7_related.tex
\section{Related Work}
\textbf{Software watermarking} has two types: static and dynamic \cite{dey2019software}. Static watermarking hides the watermark in the executable file \cite{mpanti2016graph}. Collberg \etal embedded the watermark by performing semantic equivalence transformation and reordering instructions \cite{collberg2005software}. Chen \etal used Java reflection features to find out the corresponding Java method name of the watermark bit \cite{chen2018software}. However, static watermarking requires the executable file to be obfuscated or encrypted to prevent attackers from removing the watermark. Moreover, it is vulnerable to equivalent transformation attacks \cite{ma2019xmark}. Our method generates natural watermarked variables to preserve natural and operational semantics. By using variable context graphs, \toolname offers high resilience to attacks on code blocks and variable attributes. 

In dynamic software watermarking, the watermark is stored in the program execution state \cite{wang2018exception, tian2015software}. Ma \etal proposed to transform selected conditional constructs with a control flow obfuscation technique based on the Collatz conjecture \cite{ma2019xmark}. Ren \etal embedded the watermark widget in the Android app to identify it at runtime \cite{ren2014droidmarking}. However, dynamic software watermarking requires a specific operating environment at extraction which is costly. Also, the code snippets are generally simple in logic and there are not enough execution states for watermarking. 

\textbf{Natural language watermarking} embeds a watermark by changing the syntactic and semantic nature of cover texts without affecting the original meaning of the texts \cite{qiang2023natural}. Abdelnabi \etal proposed \awt, a transformer-based method that learns to embed watermarks into unobtrusive words (\eg, articles, prepositions, and conjunctions) \cite{abdelnabi2021adversarial}. Yang \etal proposed a method \cals which uses a pre-trained language model BERT to generate the candidates \cite{yang2022tracing} for synonym substitution. Compared with WordNet, BERT generates the candidates directly based on the context of the target word. However, the source code is highly structured with strict lexical and syntax requirements. Modifications to most tokens other than variable names in code snippets will change the operational semantics.

\textbf{Naturalness of source code}. Hindle \etal are the pioneers in the field of code naturalness \cite{hindle2016naturalness}. Ray \etal leveraged cross entropy to quantify how natural the code is and suggested buggy code is less natural than bug-free code \cite{ray2016naturalness}. We also use this metric in the experiment to evaluate the watermarked code and find that the entropy of the watermarked code increases slightly compared to the non-watermarked. Naturalness loss is also used to train the pre-trained models of NatGen  \cite{chakraborty2022natgen}, which ingests unnatural code and generates natural one.

\textbf{Code variables} are crucial for the readability and comprehensibility of source code \cite{lawrie2006s, feitelson2020developers}. Various tasks aim to automatically suggest clear and meaningful variables, such as code refactoring \cite{allamanis2014learning, bavishi2018context2name, liu2019learning} and reverse engineering \cite{jaffe2018meaningful, lacomis2019dire}. While other works rename variables to create adversarial examples \cite{zhang2020generating}. Yang \etal proposed \textit{ALERT} that uses a pre-trained model to generate natural substitute variables in search of adversarial examples \cite{yang2022natural}. To reduce misleading variables, Gao \etal proposed a framework based on counterfactual reasoning that captures misleading information of variables explicitly \cite{gao2023two}. To learn semantic representations of variables, Chen \etal presented a contrastive learning method \textit{VarCLR} \cite{chen2022varclr}. Unlike previous works, we focus on how to hide a watermark on variables to track the source of code snippets. \toolname generates inconspicuous watermarked variables by learning variable naming from the generic representation of the pre-trained model on source code.

%% file: sections/3_approach.tex
\begin{figure*}[!tb]
    \centerline{\includegraphics[scale=0.6]{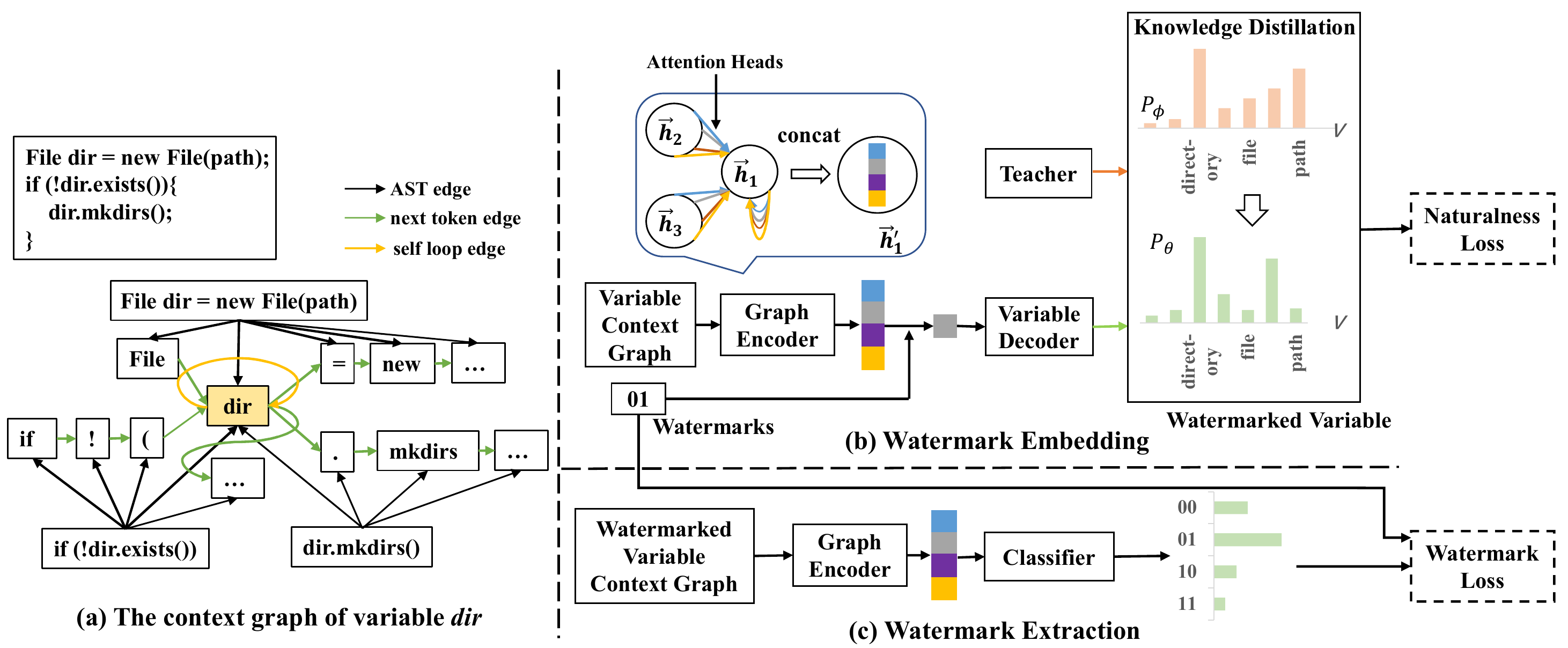}}
    \caption{The architecture of \toolname consisting of a watermark embedding and a watermark extraction module.}
    \label{fig:method}
\end{figure*}

\section{Methodology}

In this section, we will first illustrate our design goals, give an overview of \toolname following the goals, and then introduce the detail of each component of the system.

\subsection{Design Goals}
We draw insights from the digital watermarking literature for images and texts to specify our goals in designing the watermarking system for source code. The main requirements defined in \cite{cox2007digital} include: successful watermark embedding and verification, perceptual similarity, robustness to removal attacks, and security to unauthorized detection. Since program souce code has unique features different from images or texts, we set the following design goals for \toolname:

\textbf{Effectiveness:} The watermarked code should keep its utility as much as its original version. Due to the bimodal, dual-channel nature of code \cite{casalnuovo2020theory}, the utility in both channels should be preserved. On one hand, watermarking is required not to alter the well-defined operational semantics in the formal channel. Otherwise, it would risk impairing code functionality. On the other, the natural channel containing variable names, function names, etc., should be transformed with caution not to degrade the readability of the code.

\textbf{Secrecy:} The watermark should remain stealthy in the code, indistinguishable from the non-watermarked code at best. This is to prevent an adversary from detecting the watermarking pattern which is used to fake the watermark on other source code.

\textbf{Robustness:} The watermarked code should be resilient to removal attacks. Robustness prevents an adversary from removing the watermark and re-distributing the de-watermarked code without author consent. Typical removal attempts include semantic-preserving transformations, variable replacement, etc.

\textbf{Capacity and accuracy:} As a unique identifier, the watermark is expected to be sufficiently long for the embedding code. For example, if 4 bits can be embedded into a code snippet of 100 tokens, the watermark has a higher capacity than the case where only 2 bits can be embedded into the same piece of code. The number of unique watermark patterns is quadrupled in the former case ($2^4$ v.s. $2^2$), and thus allows a more diversified representation.

Following the design goals, we illustrate our design choices:

\textit{1) Embedding watermarks into the natural channel.} Since watermarking embeds 0,1 bit strings into the code, it would unavoidably change either the operational or natural semantics of the code. According to the \textit{effectiveness} requirement, we choose not to touch the formal channel, but to modify the natural channel for watermarking, i.e., to change the variable names, etc. Meanwhile, the naturalness of the code should be minimally affected to keep the original meaning as much as possible. The choice is reflected by results to RQ1 and RQ2 in Sec.~\ref{sec:result}.

\textit{2) Using a hiding network for embedding.} We partly adopt the machine translation technique to generate code with watermarks. As the deep neural network is highly expressive in representing complicated features, we choose a hiding network to embed watermarks. It is expected to acquire more diversified, natural patterns than rule-based substitution thereby evading detection, meeting the \textit{secrecy} goal. Results reflecting the choice are on RQ2 in Sec.~\ref{sec:result}.

\textit{3) Embedding bit strings to variable names.} We take one step further from design choice 1) to target at variable names for watermarking. This is because the space of variable names  is almost infinite, which admits plenty of patterns (\textit{secrecy}) without deviating from the original operational semantics (\textit{effectiveness}) . More importantly, the detection or removal of watermark bits on one variable does not affect others as variables are independently embedded, working towards the goal of \textit{robustness}. The choice is reflected by the results to RQ1, RQ2, RQ5 in Sec.~\ref{sec:result}.

\textit{4) Embedding bit strings according to variable context.} Through our preliminary tests, we found embedding bit strings to variables alone often results in poor \textit{accuracy and capacity}. The hiding network often learns a `shortcut' mapping from bit strings to variable substitution pairs, yielding rigid watermark patterns. Considered as a major contribution of this work, variable context is introduced into our watermark framework, which significantly distinguishes our method from simple variable substitution. We embed bit strings to the local context of a variable by feeding into the hiding network the graph structure surrounding the targeted variable. Hence for the same variable substitution pair, different bit strings can be embedded as the variable may be substituted in varied contexts, and thereby improving watermark accuracy and capacity. The design choice well captures the distinct feature of code, making \toolname fundamentally different from natural language watermarking. Relevant results are reported on RQ3 in Sec.~\ref{sec:result}.

\subsection{System Overview}
\toolname is a tool designed for embedding watermarks within the natural semantic space of code variables. As illustrated in Fig. \ref{fig:method}, \toolname is composed of two main components: the watermark embedding and the watermark extraction modules. As discussed in the design choices, the embedding module takes as inputs the contextual graph of each variable and a bit string, while outputing the watermarked variable to replace the original. To ensure the watermarked variables look natural in the original context, we leaverage a pre-trained teacher model to guide the generation of the watermarked variable. To extract from the watermarked code, the extraction module takes in each variable's contextual graph similar to that of embedding, and outputs the contained bit string.

All modules of \toolname need to be trained in an end-to-end manner before deployment. We define a watermark reconstruction loss $ L_{wa} $ and a naturalness loss $L_{na}$, the detail of which will be introduced shortly, to constitute the total loss function:
\begin{equation}\label{eq:loss}
	L_{t} = \alpha L_{wa} + (1 - \alpha) L_{na}
\end{equation}
with a weight parameter $\alpha$. The two losses representing watermark accuracy and code naturalness, respectively. We will analyze their competing relation closely in Sec.~\ref{sec:tradeoff}.

At the deployment stage, we first build variable context graphs for all variables in the given code snippet (with around 100 tokens on average), according to their appearance order. This is in accords with the variable naming convention, i.e., the previously named variables are often referred to by the most recent one. The variable context graphs are then fed into the embedding module of \toolname to generate watermarked variables. To prevent generating duplicated or illegal variable names, beam search is used to filter out valid ones by rules. The watermarked code is then released.

We will introduce each component of the system in the following.

\subsection{Variable Context Graph Construction}
Following design choice 3), we regard variable names as basic units for watermarking. The question is how many variable names we should use to embed one bit string. If one variable is used for embedding, the generated pattern for a single variable is quite limited due to restricted intake, resulting in unnatural watermarked variables incoherent to the context. If correlated variables across the entire snippet are chosen, the watermarked code would be vulnerable to substitution attacks, i.e., a single change of a variable would destroy the watermark. Hence we take a novel approach to take one variable as a basic unit but the variable aggregates its local context information to reflect the bit string. The idea resembles node embedding in graphs, where the embedding of a node not only captures its own feature, but also describes the local structure (one-hop neighbors, two-hop neighbors, node degrees, etc.) of the node. Likewise, the embedding takes the contextual information into account, and such structural information is a salient feature to code, i.e., Abstract Syntax Tree (AST), data flow.
 
To construct the variable context graph, we first get the AST of the code snippet. Each variable in the code is processed sequentially. For each variable, we extract its corresponding AST subtrees and merge these subtrees by combining the nodes of the target variable into one. Taking Fig.~\ref{fig:method}(a) as an example, the variable \texttt{dir} has three corresponding statements and thus three AST subtrees. The variable context graph contains three types of edges: AST edge, next-token, and self-loop edge  \cite{zhou2019devign, allamanis2017learning, wang2021code}. AST edges inherently reflect the code structure which is closely related to operational semantics; the next-token edge expresses the spatial closeness and sequential information of the code; and the self-loop edge is added to each variable node to prevent information loss of the node itself after multiple rounds of neighbor aggregation. For each edge of all types, an additional backward edge is complemented, expressed by a transposed adjacency matrix. Backward edges facilitate propagating information across the graph neural network and enhance the model's expressiveness \cite{allamanis2017learning}.

\subsection{Watermark Embedding and Extraction}
The watermark \textbf{embedding} module comprises a graph encoder and a variable decoder. Specifically, the graph encoder takes the variable context graph as the input and generates node representations through a multi-head attention network. We embed a fixed number of bits into each variable so that the watermark acts as a tunable knob in selecting one of the heads. The node representation of the selected head is fed into the variable decoder to predict which substitute to use from the token set. The detailed procedure of embedding is as follows.

The graph encoder contains several graph attention network (GAT) layers \cite{brody2021attentive,velivckovic2017graph}. The input to a GAT layer is a set of hidden features, $\boldsymbol{h} = \{ \vec{h}_{1}, \vec{h}_{2}, ..., \vec{h}_{N} \}$, $ \vec{h}_{i} \in \mathbb{R}^{F}$, where $N$ is the number of nodes and $F$ is the feature dimension of each node. The GAT layer produces a new set of node features $\boldsymbol{h}^{'} = \{ \vec{h}_{1}^{'}, \vec{h}_{2}^{'}, ..., \vec{h}_{N}^{'} \}$, $ \vec{h}_{i}^{'} \in \mathbb{R}^{{F}^{'}}$ as its output. Generally, if the bit length of the watermark is $L$, we set $K = 2^{L}$ for the $K$-head attention \cite{vaswani2017attention}. Each head independently executes the attention mechanism, and their features are concatenated in the following output feature representation: 
\begin{equation}
\vec{h}_{i}^{'} = \mathop{||}\limits_{k=1}^K \sigma ( \sum\limits_{j \in N_{i}} \alpha^{k}_{ij} \boldsymbol{W}^{k} \vec{h}_{j}), 
\end{equation}
where $||$ represents concatenation, $\sigma$ is an activation function, $N_{i}$ is neighborhood of node $i$ in the graph, $\alpha^{k}_{ij}$ are the normalized attention coefficients computed by the $k$-th attention mechanism ($\alpha^{k}$), and $\boldsymbol{W}^{k}$ is the linear layer's weight matrix. Fig.~\ref{fig:method} (b) shows an illustrative example of $2$-bit watermark embedded with $4$-head attention for Node \texttt{dir}. Different colors denote independent attention mechanisms. The aggregated features from each head are concatenated to obtain $\vec{h}_{i}^{'}$. The second block (a gray square) of the node representation is selected by the watermark \texttt{01}.

The variable decoder of the embedding module is implemented by a Long Short-Term Memory (LSTM) network \cite{graves2012long}, which generates a substitute variable for the original one. To generate variables looking more `natural,'  we adopt CodeBERT \cite{feng2020codebert}, a bimodal pre-trained model for programming language and natural language, as a teacher model to distill the knowledge to our variable decoder. Fed with the statement containing the target variable, the teacher network generates a probabilistic vector for the candidate tokens. Such a probabilistic vector of the CodeBERT, capturing the semantic connection between natural language and programming language, serves as a general-purpose representation to guide the training of the variable decoder. In practice, we average out the probabilistic vectors over multiple statements concerning the target variable. Only the top-$k$ scored tokens are considered as relevant and we normalize their probabilities by applying a softmax function. The resulting probabilistic vector is used as the soft label to help the variable decoder generate more natural variables. The naturalness loss is defined as the cross entropy \cite{de2005tutorial} between CodeBERT's and variable decoder's outputs:
\begin{equation}
    L_{na} = - \sum\limits_{t} \sum\limits_{w \in V} \left[ P_{\phi} (y_{t} = w) \cdot \log P_{\theta} (y_{t}=w) \right],
\end{equation}
where $y_t$ is the $t$-th subtoken for the predicted variable, $w$ is a word from the vocabulary set $V$. $P_{\phi} (\cdot)$ is the soft label estimated by CodeBERT with parameters $\phi$ and $P_{\theta} (\cdot)$ is the output of the student model $\theta$. Note that $\phi$ is fixed during the distillation process and \toolname only uses CodeBERT in the training phase. An illustration of distillation is given in Fig.~\ref{fig:method} (b).

The \textbf{extraction} module comprises a graph encoder and a classifier. Specifically, the graph encoder takes the variable context graph as input and generates node representations. Subsequently, the variable node representation is fed into the classifier to predict the watermark. The graph encoder shares the same structure with that of the embedding module, whereas the classifier is a two-layer perceptron. Since we treat the watermark extraction as a multi-classification problem, the watermark loss is defined as 
\begin{equation}
L_{wa} = - \sum\limits_{c \in C} y_{c}\log(p_{c}),
\end{equation}
where $y_c$ is the label of class $c$ from $|C|=2^{L}$ categories, and $p_c$ is the prediction. The extraction module is depicted in Fig.~\ref{fig:method} (c).

In the end-to-end training to minimize the total loss $L_t$ (Eq.~\eqref{eq:loss}), to facilitate backpropagation from the extraction module, our embedding module uses a Gumbel Softmax approximation with one-hot encoding in the forward pass, allowing differentiation in the backward pass \cite{jang2016categorical, kusner2016gans}.

%% file: sections/4_experiment_set.tex
\section{Experimental Setup}
We provide the detail of the experimental design.

\begin{table}[!tb]
    \caption{Statistics of Datasets. * The average number of tokens for a function is 99.06.}
    \begin{center}
        \begin{tabular}{lcccc}
            \toprule
            Dataset (Description)   & Train   & Valid  & Test   & Unit     \\
            \midrule
            D0 (CodeSearchNet-Java) & 424,451 & 15,328 & 26,909 & Function \\
            D1 (Filtered from D0)   & 164,923 & 5,183  & 10,955* & Function \\
            D2 (Graphs of D1)       & 542,227 & 15,337 & 37,268 & Graph    \\
            \bottomrule
        \end{tabular}
        \label{table:dataset}
    \end{center}
\end{table}
\textbf{Dataset and pre-processing.}
We use the Java dataset of CodeSearchNet \cite{husain2019codesearchnet} which contains \textless comment, code \textgreater  pairs collected from non-fork open source projects, where \textit{code} means the code snippet of a function and \textit{comment} is the code description mostly from Javadocs.

We preprocess the dataset by steps. The raw data in CodeSearchNet is referred to as {D0}. 
\blueDelete{To ensure the corresponding comments are relevant to the code, we apply filtering rules from CodeXGLUE to remove unrelated comments such as URL links and HTML tags.} \blue{To measure natural semantics, we need the comments are relevant to the code, so we apply filtering rules from CodeXGLUE \cite{lu2021codexglue} to remove unrelated comments.} The filtered dataset is {D1}. Then we utilize the tree-sitter parser\footnote{https://tree-sitter.github.io} to parse the abstract syntax tree of the code in D1 and construct a graph for each variable. The collected graphs compose D2. Since the training data requires a probability vector assigned by CodeBERT, we only take variables in the first 510 tokens of each function as the training data. The division of training/validation/test examples are consistent across {D0}, {D1} and {D2}. Statistics of the pre-processed datasets are provided in Table \ref{table:dataset}.

\textbf{Baselines.} Due to a lack of baselines in code watermarking, we select two natural language watermarking tools --- \awt \cite{abdelnabi2021adversarial} and \cals \cite{yang2022tracing}, and build their variants \awtcode and \calscode as comparison. \awt is composed by a sequence-to-sequence Transformer trained on WikiText-2, and we faithfully follow the original implementation. In contrast, \awtcode shares a similar structure but is trained on {D1} to adapt the model to source code. \cals employs a sequential incremental watermarking scheme through context-aware lexical substitution. We follow their original implementation using \textit{bert-base-cased}\footnote{\url{https://huggingface.co/bert-base-cased}} for candidate generation and \textit{roberta-large-mnli}\footnote{\url{https://huggingface.co/roberta-large-mnli}} for similarity score calculation. In \calscode, the candidate generation model is replaced as \textit{codebert-base-mlm}\footnote{\url{https://huggingface.co/microsoft/codebert-base-mlm}} to be coherent with code.

\textbf{Evaluation metrics.} We evaluate \toolname from four perspectives: 1) operational semantics, 2) natural semantics, 3) watermark performance and 4) robustness. For 1), we adopt syntax analysis as a necessary condition for operational semantic equivalence, as other methods such as software testing, formal verification require custom rules or test cases. The watermarked code are checked by its abstract syntax tree for errors (AST check). Since watermarking is likely to modify letter cases for keywords and turn them illegal, we also apply a list of java keywords\footnote{\url{https://github.com/soarsmu/attack-pretrain-models-of-code/blob/main/python_parser/run_parser.py\#L24}}, including reserved words and commonly used class names such as \texttt{String}, to check the illegal changes by regular expressions (keyword check).

\blueDelete{For 2), we evaluate natural semantics by direct evaluation metric N-gram entropy, as well as the performance of watermarked code on tasks including code summarization and code search.}

\blue{For 2), we employ several metrics, namely BLEU, MRR, Entropy, and VarSim, to assess the natural semantics of watermarked code.  BLEU and MRR are metrics utilized for code search and code summary tasks, respectively, with both tasks focusing on the semantic relationship between code and natural language. Following CodeXGLUE \cite{lu2021codexglue}, we fine-tune two CodeBERT models on code summary and code search tasks to test watermarked code, respectively. Entropy, previously used in research to differentiate between the naturalness of buggy and bug-free code \cite{hindle2016naturalness, jiang2022bugs}, is computed using the 3-gram model provided by \cite{hindle2016naturalness}. Lastly,  VarSim is calculated using the state-of-the-art method proposed by \cite{chen2022varclr}, which effectively measures the similarity between two variables. In general, lower entropy, higher BLEU, higher MRR, and higher VarSim indicate that the watermarked code is more natural.}

\blueDelete{As to 3), we consider the conventional performance metrics for watermarks, i.e., bitwise accuracy, capacity and efficiency. }

\blue{As to 3), we mainly consider the following aspects for the watermarking performance: \emph{accuracy}, \emph{capacity} and \emph{efficiency}, measured respectively by bitwise accuracy (BitAcc), bits per token (BPT) and average time to embed and extract the watermark. BitAcc is the percentage of bits that are correctly extracted. Random guessing would result in 0.5. BPT is the average number of bits embedded in each token. Notably, how many bits could be embedded per token actually depends on the specific methods. For example, \awt embeds a fixed number of four bits per function whereas \cals relies on the number of tokens passing the synchronization test. \toolname depends on the number of variables in the code. Embed time and extract time for each model are measured on the same machine. Each model processes one input sample at a time (i.e., batch size is 1) and the total time is averaged over all samples. Note that the neural network model would be pre-loaded to the GPU before the test, and the time for loading the model is not included. In general, higher bitwise accuracy, higher BPT, and lower time indicate a high-performance watermarking method.}

For 4) robustness, we evaluate the watermark extraction accuracy in face of different de-watermarking attacks. Three levels of attacks are adopted to test the resilience of each method.

\textbf{Implementation.} Our implementation is based on PyTorch. Both the embedding and extracting networks consist of 2-layer GATs with 4 attention heads, as provided by the Deep Graph Library\footnote{https://docs.dgl.ai/}. All hidden layers and word embeddings have a dimensionality of 512. We use the Adam optimizer \cite{kingma2014adam} with a learning rate of 0.00025. A Gumbel temperature of 0.5 is used \cite{shetty2017speaking, shetty2018a4nt}, along with a dropout probability of 0.1 and ReLU activations. Each variable is assigned with 2-bit watermarks. The value of $\alpha$ in Eq. \ref{eq:loss} is set to 0.6 and will be discussed in Sec. \ref{sec:tradeoff}. We train and validate \toolname on the training and validation sets of D2 and use the test set of D1 to report the performance. We run all experiments five times and report the average results. All experiments are conducted on an Intel(R) Xeon(R) Gold 6240C CPU @ 2.60GHz Linux 20.04 server with 256GB RAM and four NVIDIA GeForce RTX 3090 graphic cards.

%% file: sections/5_results.tex
\section{Experiment Results and Analysis}\label{sec:result}
We aim to answer the following research questions with our experiments:
\begin{itemize}
	\item \textbf{RQ1 (Operational Semantics).} How does watermarking affect the operational semantics of code?
	\item \textbf{RQ2 (Natural Semantics).} How does watermarking affect the natural semantics of code?
	\item \textbf{RQ3 (Watermark Performance).} What are the watermark accuracy, watermark capacity, and time overhead of watermarking methods?
	\item \textbf{RQ4 (Trade-off).} How does \toolname balance the watermark accuracy with the natural semantics of code?
	\item \textbf{RQ5 (Robustness).} How robust is \toolname?
\end{itemize}

\begin{figure}[!tb]
  \centerline{\includegraphics[scale=0.9]{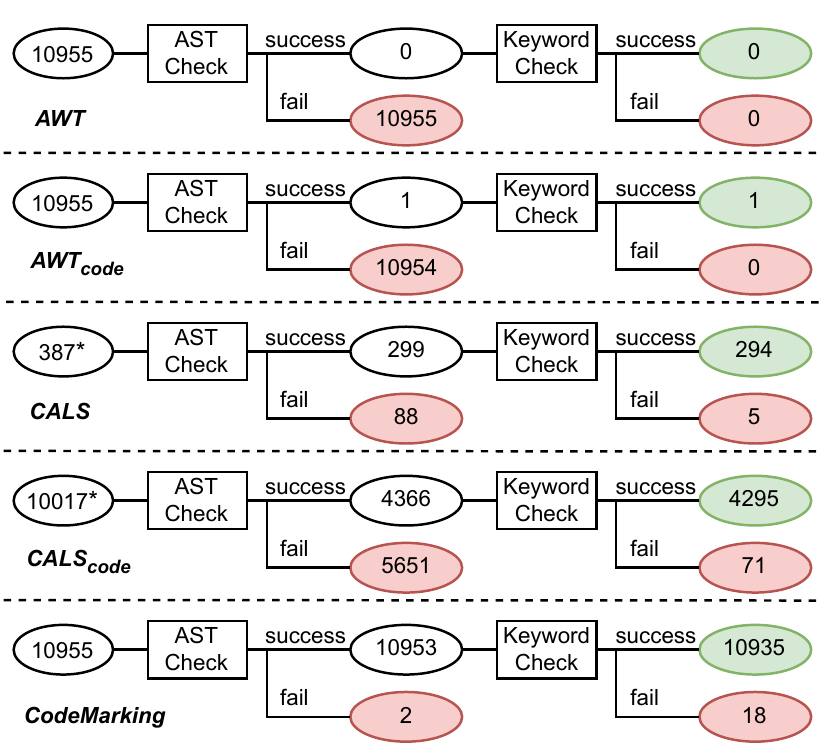}}
  \caption{Comparison of operational semantics changes of various watermarking methods. * Of the 10955 functions, \cals and \calscode successfully embed watermarks for 387 and 10017 functions, respectively.}
  \label{fig:operational-semantics}
\end{figure}

\subsection{Operational Semantics (RQ1)}
Fig.~\ref{fig:operational-semantics} shows how the baseline methods and \toolname preserve the operational semantics after watermarking, especifically the proportion of code that have passed and failed AST and keyword checks. \toolname modifies only variable names, which means that 99.81\% of the samples still work the same way as before. Only 2 samples failed the AST check and 18 samples failed the keyword check. We manually reviewed the original code for the 20 samples and found that they had issues already presented in the test set. 

\begin{figure}[!tb]
  \centering
  \includegraphics[width=0.9\linewidth]{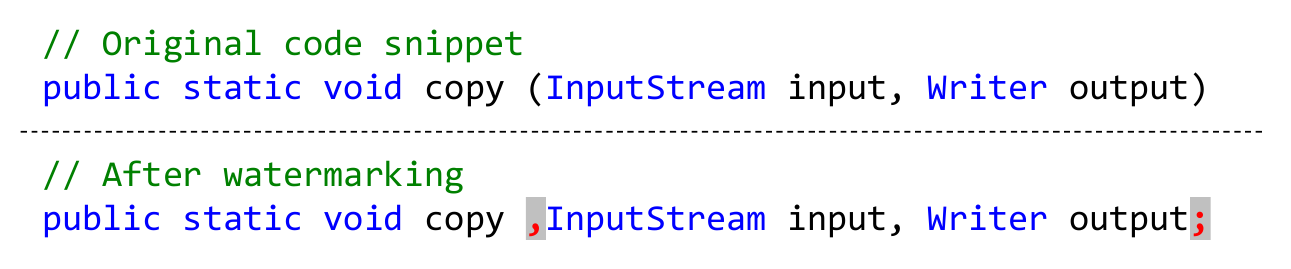}
  \caption{A failure case of \awtcode. The code snippet before (top) and after (bottom) watermarking. The place highlighted in red violates the operational semantics.}
  \label{fig:awt-grammar-fail-case}
\end{figure}

Both \awt and \awtcode suffer major failures in the AST check. Few, if any, watermarked code generated by \awt and \awtcode pass the AST check, which means the watermarked code produced by AWT models are very likely to contain grammatical errors. One possible interpretation is that AWT uses a generative model to directly generate the watermarked code. Even though AWT is trained with a reconstruction loss that heuristically guides the model to ``reconstruct'' the input code, no explicit syntax constraint is enforced on the model. Therefore, the model may still make unrestricted changes to the inputs and produce syntactically incorrect results. Fig.~\ref{fig:awt-grammar-fail-case} provides an illustrating exmaple for the failure case of \awtcode. While \awtcode successfully keeps all keywords and variable names intact, it mistakenly changes the parentheses in the function signature into a comma and a semicolon, which violates the grammar specifications of Java and thus leads to a syntax error.

\begin{figure}[!tb]
  \centering
  \subfloat[]{\includegraphics[width=0.6\linewidth]{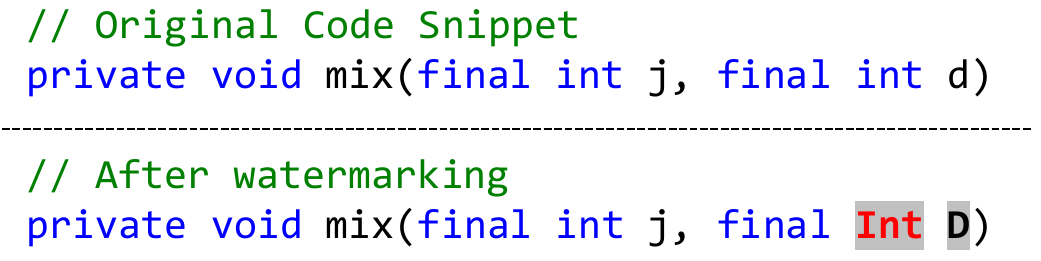}}\quad
  \subfloat[]{\includegraphics[width=0.6\linewidth]{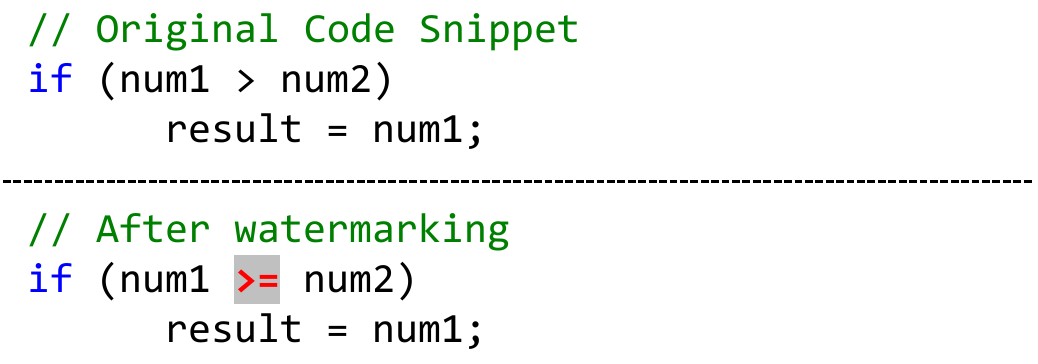}}
  \caption{Two failure cases of \calscode. The code snippet before (top) and after (bottom) watermarking. The place highlighted in red changes the operational semantics, while highlighted in black preserves the operational semantics.}
  \label{fig:calscode-grammar-fail-case}
\end{figure}

\blue{\cals successfully watermarked only 387 (3.53\%) code snippets. This low success rate can be attributed to its focus on natural language texts, which makes it difficult to generate suitable and similar candidate tokens within the code context. To address this limitation, we implemented \calscode, which increased the number of embedded watermarks to 10,017 (91.44\%) by utilizing CodeBERT as the candidate generation model. Fig.~\ref{fig:operational-semantics} illustrates that approximately 40\% of the watermark code generated by \calscode passed the AST and keyword check.}

\blue{However, our manual inspection revealed that a significant proportion of the watermarked code generated by \calscode, even after passing the AST and keyword checks, could still change the operational semantics. This issue arises because the watermarked tokens generated by CodeBERT may include both uppercase and lowercase forms of the same token. Consequently, token pairs such as \texttt{int} and \texttt{Int} change the operational semantics. While changing from \texttt{d} to \texttt{D} is allowed, as demonstrated in Fig.~\ref{fig:calscode-grammar-fail-case}(a). Various symbols with high similarity scores can introduce logic errors, as shown in Fig.~\ref{fig:calscode-grammar-fail-case}(b).}

\begin{center}
  \begin{tcolorbox}[colback=gray!10,
      colframe=black,
      arc=1mm, auto outer arc,
      boxrule=0.5pt,
    ]
    Answers to RQ1: \toolname is the most effective in preserving the operational semantics of the code, while \awt is the least effective.
  \end{tcolorbox}
\end{center}

\subsection{Natural Semantics (RQ2)}

\begin{table}[!tb]
  \caption{Comparison of natural semantics changes of various watermarking methods.}
  \label{table:naturalness}
  \centering
  \begin{tabular}{@{}cccc@{}}
    \toprule
    Methods         & BLEU  & MRR    & Entropy \\ \midrule
    Non-watermarked & 17.15 & 0.8180 & 6.6943  \\ \midrule
    \awt            & 10.77 & 0.3593 & 5.8702  \\
    \awtcode        & 15.54 & 0.7796 & 6.8932  \\ \midrule
    \cals           & 17.29 & 0.8224 & 6.7691  \\
    \calscode       & 17.13 & 0.8190 & 6.7940  \\ \midrule
    \toolname       & 15.74 & 0.8064 & 6.7684  \\ \bottomrule
    \end{tabular}
  \end{table}


Table \ref{table:naturalness} summarizes the results of naturalness. Our results indicate that the BLEU metric is more sensitive to natural semantic changes in code watermarking, whereas MRR and entropy are not. BLEU is sensitive because it summarizes code comments based on key tokens. \blueDelete{Among the three models used to test BLEU, $M_C$ leads to the most significant drop in BLEU after watermarking compared to before, and thus can be considered as the worst case. Therefore, we use the BLEU under $M_C$ as the metric in subsequent experiments.} Entropy, on the other hand, is not effective as it is likely to be insensitive to small changes in the code. Earlier studies also have arrived at similar conclusions, suggesting that although code containing bugs may exhibit higher entropy than code without bugs, fixing the bugs does not significantly decrease entropy or enhance naturalness \cite{jiang2022bugs}.

Both \awt and \awtcode have losses in natural semantics, with \awt having a more significant drop. For \awtcode, BLEU drops by around 10\% and MRR drops by 5\%, while for the original version of \awt, BLEU drops by over 30\% and MRR drops by over 50\%. The result is not surprising, as \awt is trained on a natural language dataset while \awtcode is trained on a dedicated source code dataset. Notably, entropy for code produced by \awt decreases significantly. While lower entropy usually means higher naturalness, it is not the case for \awt. After manually inspecting the outputs, we find that the reason for the low entropy of \awt is not because of increased naturalness, but due to frequently-appearing repeated tokens. \awt has a failure pattern where it outputs the same token repeatedly (e.g., producing a sequence of ``the, the, \dots, the''). In the original dataset, no sample contains any single token (excluding the special ``unk'' token) with a proportion over 30\%, but in the watermarked output of \awt, there are 1048 such samples, which accounts for almost 10\% of the test set. Such repeated tokens significantly reduce the entropy of the output, but do not contribute to naturalness.

Since the number of successfully embedded watermark samples by \cals is 294, which is too small, the comparison of its naturalness index has no reference value. All the changes in \calscode's three naturalness indicators are small and within the margin of error. One possible explanation is that for \calscode, the embedded watermark is partially situated on the symbols, and the replacement words not only guarantee a high SR score before and after the substitution, but the majority of the replacement words exhibit differences solely in terms of capitalization, singular/plural forms and tense based on our observations, resulting in a minimal impact on the naturalness.

\begin{table}[!tb]
  \centering
  \caption{Number of watermarked code snippets that pass the grammar check and exceed naturalness metric thresholds. \awt and \awtcode are omitted as only 0 and 1 code snippets pass the grammar check, respectively.}
  \label{table:naturalness-thresholding}
  \begin{tabular}{@{}cccccc@{}}
    \toprule
    Method    & BLEU     & MRR       & Entropy & VarSim (Mod) \\
              & $\ge$ 17 & $\ge$ 0.8 & $\le$ 7 & $\ge$ 0.55           \\ \midrule
    \cals     & 53       & 11        & 63      & 2                \\
    \calscode & 1361     & 231       & 1872    & 226              \\
    \toolname & 4045     & 713       & 5948    & 1840             \\ \bottomrule
    \end{tabular}
\end{table}

It is worth mentioning that the results in Table~\ref{table:naturalness} are evaluated on the entire test set, and do not take syntactic correctness into consideration. To further investigate the naturalness of \emph{syntactically correct} watermarked code, we only retain watermarked code that are syntactically correct, and count the number of such code that meet certain thresholds for natural semantic metrics (namely BLEU, MRR, Entropy, and VarSim). Thresholds for BLEU, MRR and entropy are selected according to the scores of non-watermarked code on downstream tasks, and the threshold for VarSim is selected according to the case study in section~\ref{sec:tradeoff}. 

Table~\ref{table:naturalness-thresholding} reports the thresholds and the results. Note that we omit the results of \awt and \awtcode because almost all watermarked code generated by AWT models fail the grammar check. For all four metrics, \toolname stands out with the most number of syntactically correct watermarked code that meets the thresholds, which indicates that \toolname could successfully watermark most of the code while guaranteeing both operational and natural semantics.

\begin{center}
  \begin{tcolorbox}[colback=gray!10,
      colframe=black,
      arc=1mm, auto outer arc,
      boxrule=0.5pt,
    ] 
    Answers to RQ2: Although \toolname does not score the highest (\calscode) by the average naturalness performance, it successfully watermarks most of the code while guaranteeing operational and natural semantics. 
  \end{tcolorbox}
\end{center}

\begin{table}[!tb]
  \caption{Watermark performance of various watermarking methods. BPT is the abbreviation for bits per token and the unit of time is seconds.}
  \begin{center}
    \begin{tabular}{lcccc}
      \toprule
      Methods   & BitAcc & BPT    & Embed Time & Extract Time \\
      \midrule
      \awt      & 0.9556 & 0.0639 & 0.0307     & 0.0018       \\
      \awtcode  & 0.9301 & 0.0639 & 0.0363     & 0.0019       \\
      \midrule
      \cals     & 0.9956 & 0.0161 & 14.9524    & 14.9329      \\
      \calscode & 0.7046 & 0.0494 & 25.3142    & 24.8555      \\
      \midrule
      \toolname & 0.8745 & 0.0826 & 0.0934     & 0.0317       \\
      \bottomrule
    \end{tabular}
    \label{table:watermark}
  \end{center}
\end{table}

\subsection{Watermark Performance (RQ3)}

\vspace{3mm}
From Table~\ref{table:watermark}, we can tell \awt and \awtcode perform similarly, but \awt's embedded code is impractical because it does not pass grammatical analysis. \cals enjoys a high accuracy but a limited capacity, while \calscode sacrifices accuracy to increase capacity. The watermark accuracy of \toolname is lower than that of both \awt and \cals, but still within an acceptable range at 0.87. Notably, \toolname outperforms both \awt and \cals in watermark capacity by 29.3\% and 67.2\%, respectively. A higher capacity means a longer bit string can be embeded as watermark, helping to effectively identify the code. W.r.t. efficiency, \cals and \calscode take much longer than other methods due to their sequential watermarking scheme. For a code snippet with $n$ tokens, they require the pre-trained model to generate candidates at least $8n$ times. Additionally, the watermarking process for different tokens in a single code snippet is done in a serial manner, making running time optimization difficult. In contrast, \toolname merely learns to name variables using knowledge distillation in training and does not involve any pre-trained model in the inference, and hence is faster. It takes less than 0.1 seconds to embed or extract the watermark, with most of the time spent building graphs.

\begin{center}
  \begin{tcolorbox}[colback=gray!10,
      colframe=black,
      arc=1mm, auto outer arc,
      boxrule=0.5pt,
    ]
    Answers to RQ3: \toolname exhibits a significantly greater watermark capacity, surpassing that of \awt and \cals by 1.29 and 5.13 times, respectively. \toolname is highly efficient in generating watermarked code, taking less than 0.1 seconds. The bit accuracy of \toolname is 0.87.
  \end{tcolorbox}
\end{center}

\begin{figure}[!tb]
  \centerline{\includegraphics[scale=0.68]{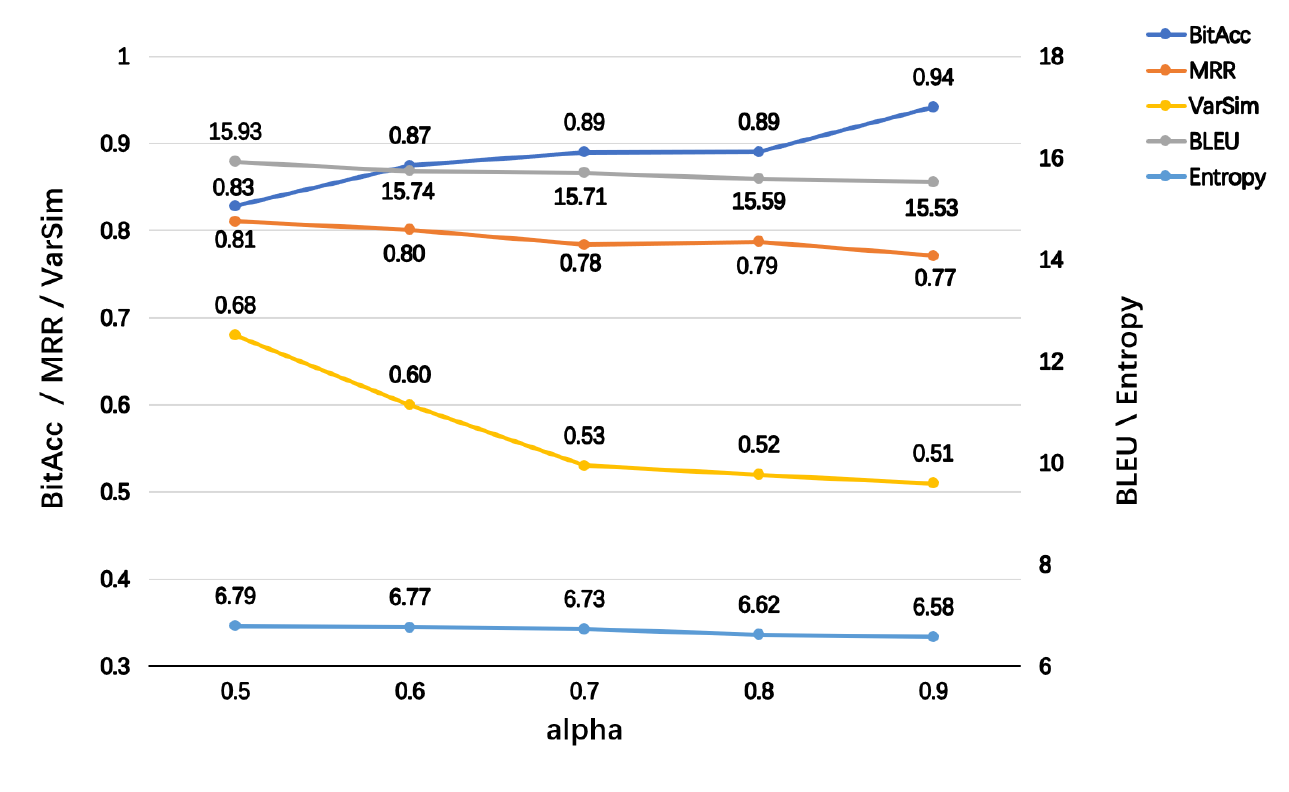}}
  \caption{\toolname: trade-off between code naturalness and watermark accuracy.}
  \label{fig:trade-off}
\end{figure}

\begin{table*}[!tb]
  \caption{Case study on variable similarity between original variables and watermarked variables.}
  \begin{center}
    \begin{tabular}{ccc|ccc|ccc}
      \hline
      \makecell[c]{Original                                                                                          \\ Var}   & \makecell[c]{Watermarked \\ Var} & \makecell[c]{VarSim}   & \makecell[c]{Original \\ Var}   & \makecell[c]{Watermarked \\ Var} & \makecell[c]{VarSim} & \makecell[c]{Original \\ Var}   & \makecell[c]{Watermarked \\ Var} & \makecell[c]{VarSim} \\
      \hline
      userDetails     & userInfo         & 0.915 & logger      & logFile  & 0.680 & batch           & token  & 0.448 \\
      a2              & a1               & 0.853 & builder     & instance & 0.555 & queue           & token  & 0.404 \\
      doc             & document         & 0.846 & webSocket   & webNode  & 0.543 & numErrorsBefore & numMMM & 0.385 \\
      iterator        & iter             & 0.800 & eventType   & thisType & 0.498 & targets         & n11    & 0.228 \\
      defaultValueMap & defaultValueInfo & 0.736 & descriptors & descMM   & 0.479 & servers         & defD   & 0.198 \\
      \hline
    \end{tabular}
    \label{table:similar-var}
  \end{center}
\end{table*}

\subsection{Watermark Accuracy v.s. Naturalness (RQ4)}
\label{sec:tradeoff}

\toolname adjusts the $\alpha$ parameter to balance the naturalness of the watermarked code and its bit accuracy for watermark. 
As shown in Fig.~\ref{fig:trade-off}, among the four metrics we used to evaluate naturalness (BLEU, MRR, Entropy, and VarSim), VarSim exhibits the most significant changes as $\alpha$ varies, which also shows the opposite trend with the bit accuracy, indicating a clear tradeoff between code naturalness and watermark accuracy. As $\alpha$ increases beyond 0.6, the improvement in accuracy is minimal, displaying a desirable tradeoff at $0.6$. Therefore, we choose $\alpha$ to be 0.6 by default in the tests.

To gain a better understanding of watermarked variables, we conduct a case study on \toolname with $\alpha$ set to 0.6, and generate results in Table~\ref{table:similar-var}. We observe that when the variable similarity is around 0.8, the watermarked variable is highly natural, possibly being a synonym or abbreviation of the original variable, such as the variables \texttt{doc} and \texttt{document}. When the variable similarity is about 0.55, the watermarked variable is relatively similar to the original variable, possibly sharing a common sub-token. A lower VarSim would result in totally irrelevant variables.

\begin{center}
  \begin{tcolorbox}[colback=gray!10,
      colframe=black,
      arc=1mm, auto outer arc,
      boxrule=0.5pt,
    ]
    Answers to RQ4: \toolname shows a clear tradeoff between watermark accuracy and naturalness by adjusting $\alpha$.
  \end{tcolorbox}
\end{center}

\begin{table}[!tb]
	\caption{Three types of attacks.}
	\label{tab:code-transforms}
	\centering
  \footnotesize
  \resizebox{\linewidth}{!}{
    \begin{tabular}{lp{0.23\linewidth}p{0.63\linewidth}}
      \toprule
      \textbf{Type}       & \textbf{Attributes}           & \textbf{Description}                                                                                                                 \\ \midrule
      \multirow{4}{*}{I}  & Increment Expr.         & Styles for increment or decrement expr.: \texttt{i++;}, \texttt{++i}, \texttt{i+=1;}, \texttt{i=i+1;}           \\
      & Loop Structures              & \texttt{for} loops or \texttt{while} loops                                                                                               \\
      & Cond. Structures       & \texttt{if} structures or \texttt{switch} structures                                                                                     \\
      & Nested Conditions            & Whether multiple \texttt{if} conditions are nested or merged: \texttt{if(a)\{if(b)\}} or \texttt{if(a \&\& b)\{\}}           \\ \midrule
      \multirow{5}{*}{II} & Naming Style        & Variable naming: \texttt{camelCase}, \texttt{PascalCase}, \texttt{snake\_case}, \texttt{\_underscore\_init}                       \\
      & Variable Def.          & Location of defining local variables: at the beginning of the function or defined on first use                                    \\
      & Variable Init.      & Location of initializing local variables: defined and initalized simultaneously or separately: \texttt{int i=0;} or \texttt{int i; i=0;} \\
      & Multiple Defs.    & Variables of the same type are defined together or separately: \texttt{int i, j;} or \texttt{int i; int j;}               \\
      & Temporaries      & Whether to introduce temporary variables for assignments: \texttt{x=i++;} or \texttt{tmp=i++; x=tmp;}                                    \\ \midrule
      III                 & Variable Names   & Random variable name substitution                                                                                                    \\ \bottomrule
    \end{tabular}
  }
\end{table}

\begin{table}[!tb]
  \caption{The robustness of \toolname to three types of attacks.}
  \begin{center}
    \begin{tabular}{lccccc}
      \toprule
      Attack Type & BitAcc & VarSim & BLEU  & MRR    & Entropy \\
      \midrule
      Non-attack     & 0.8745 & 0.60   & 15.74 & 0.8064 & 6.7684  \\
      \midrule
      I              & 0.8745 & 0.60   & 15.79 & 0.7752 & 6.7571  \\
      \midrule
      II             & 0.8085 & 0.58   & 15.69 & 0.7742 & 6.8486  \\
      \midrule
      III (Rename 25\%)     & 0.7817 & 0.55   & 15.71 & 0.7974 & 6.7970  \\
      III (Rename 50\%)     & 0.6873 & 0.51   & 15.56 & 0.7889 & 6.8433  \\
      III (Rename 75\%)     & 0.5947 & 0.46   & 15.40 & 0.7853 & 6.8933  \\
      III (Rename 100\%)    & 0.5014 & 0.42   & 15.49 & 0.7827 & 6.9071  \\
      \bottomrule
    \end{tabular}
    \label{table:attack}
  \end{center}
\end{table}

\subsection{Robustness (RQ5)}
The watermark should be resilient against various attacks to remove it. Therefore, we assume an adversary whose goal is to remove the watermark from the watermarked code while maintaining the utility, so that the code could be used or redistributed without the watermark. To maintain utility, an adversary may perform semantic-preserving code transformations on the watermarked code. Hence we reuse the semantic-preserving transformations proposed in RopGen~\cite{li2022ropgen} but only retain the attributes applied to Java functions as our dataset is Java-based. In total, there are 10 different types of attributes, which are detailed in Table~\ref{tab:code-transforms}. Since the watermark information is embedded into variable names and their local graph contexts, we further divide the attributes into three categories according to their relevance to variable names and contexts. Type I attributes are mainly related to block level syntactic structures, such as loops and conditional branches, which are the least relevant to the embedding context. Type II attributes involve changes to the local context of variables, while changing these attributes do not directly modify variable names, it may affect the graph context of variables, thus influencing the extraction of watermarks. Type III attributes are directly related to variable names, the altering of which would arguably lead to a greater loss of the watermark information.

To assess the robustness of our method under these three levels of attacks, we first perform code transformations on the watermarked code. For Type I and Type II attributes, we scan through each applicable attribute and apply random transformations to the attribute. For Type III (variable names), we consider randomly substituting 25\%, 50\%, 75\% and 100\% of the variable names in the function. We then try to extract watermarks from the transformed code to evaluate extraction bit accuracy.

\vspace{3mm}
Table \ref{table:attack} shows how well \toolname withstands different attacks. The first row displays the results without any attacks. Type I attacks mainly change code block information, but do not affect the code context graphs much, so \toolname is robust to this type of attack. Type II attacks change certain variable attributes, such as variable style and initialization method, which modifies the code context graph, but does not change most of the structure and semantics of the variable context graph. Type III attacks change variable names, which makes \toolname unable to extract watermarks accurately from the context graph using variable guidance. However, this attack causes variable names to almost completely differ from the original, leading to an unbearable loss of natural semantics --- 30\% drop in VarSim. Hence the attack is unsuccessful either.

\begin{center}
  \begin{tcolorbox}[colback=gray!10,
      colframe=black,
      arc=1mm, auto outer arc,
      boxrule=0.5pt,
    ]
    Answers to RQ5: \toolname shows high resilience to attacks on code block and variable attribute modification. Variable name alteration removes watermarks with a significant loss of naturalness, which is considered an unsuccessful attack.
  \end{tcolorbox}
\end{center}

%% file: sections/6_discussion.tex
\section{Discussion}

The \textbf{limitation} is the watermark capacity. This limitation arises because hiding watermarks in code is more challenging than that in images or text, due to more stringent requirements. Since our method focuses solely on watermarking variables, if a code snippet does not contain variables, our method would not work. Fortunately, variables are commonly used in code, especially in functions of highly complex logic. Additionally, due to the naturalness of the code, we only embed a 2-bit string into each variable, which could also limit the capacity.

\textbf{Threats to Validity.} The first threat comes from the validity of the experimental results in measuring the operational and natural semantics. In terms of operational semantics, we check if the watermarked code can parse the syntax tree without modifying Java's reserved words. However, this inspection alone is not enough. Our findings show that for \calscode,  95\% of the code that passes the operational semantics check modifies tokens other than variable, like the class or API name, which often alters the operational semantics. Unfortunately, using formal analysis or software testing tools for this purpose is too expensive.

As for natural semantics, our empirical studies rely on four models: a code search model, a code summary model, a code entropy model, and a variable similarity model. We choose these models for a few reasons. First, their implementations are open source. Second, the code search and code summary models effectively capture semantics across programming and natural languages. The code entropy model has been previously used to examine the naturalness of bugs \cite{jiang2022bugs}. Apart from that, the variable similarity model offers a more intuitive reflection of the semantic similarity between embedded watermark variables and the original variables.

Another threat to internal validity is that the raw data provided by CodeSearchNet is in the format of plain text. We parse statements based on rules provided by \cite{zhang2022diet}. However, there could be irregular patterns that the parser can not recognize. This could cause noise in our dataset and affect the results of our research. An external threat is that we have merely conducted experiments in Java. Although constructing variable context graphs is language-agnostic, other programming languages such as LISP and Erlang could have different semantic patterns to which our method/results may not generalize. 

%% file: sections/8_conclusion.tex
\section{Conclusion}

In this paper, we address the issue of tracing the provenance of code snippets. We propose a solution \toolname, a code watermarking system designed to hide bit strings in source code while preserving its operational semantics and naturalness. Our approach is novel in adopting a machine translation-like framework as well as a graph neural network to generate new variables to replace the old ones. By modeling each variable as nodes on graphs, we let each variable feature `absorb' its context as a node aggregates neighborhood information on graphs. The entire network, including an embedding and an extraction module, is trained end to end. Despite its superior performance to SOTA in Java, we wish to extend our system to a wider variety of programming languages and further improve the watermarking capacity.